# Progress and Prospects in Neutrino Astrophysics*

John N. Bahcall[§], Kenneth Lande[†], Robert E. Lanou, Jr.[‡], John G. Learned[§§],

R.G. Hamish Robertson[**], Lincoln Wolfenstein[††]

[§]Institute for Advanced Study, Princeton, NJ 08540

[†]Department of Physics, University of Pennsylvania, Philadelphia, PA 19104-6394

[‡]Department of Physics, Brown University, Providence, RI 02912

[§§]Department of Physics & Astronomy, University of Hawaii, Honolulu, HI 96822

[**]Physics Division, Los Alamos National Laboratory, Los Alamos, NM 87545 and
Department of Physics, University of Washington, Seattle, WA 98195

[††]Department of Physics, Carnegie Mellon University, Pittsburgh, PA 15213

The neutrino deficits observed in four solar neutrino experiments, relative to the theoretical predictions, have led to fresh insights into neutrino and solar physics. Neutrino emission from distant, energetic astronomical systems may form the basis for a new field of astronomy. Additional experiments are needed to test explanations of the solar neutrino deficits and to allow the detection of more distant sources.

## Why neutrino astrophysics?

Neutrino astrophysics has developed from theoretical calculations to an experimental discipline over the last three decades. As we shall see in later sections of this paper, solar neutrinos have been observed in four underground experiments and a burst of neutrinos was detected from a supernova in the Large Magellanic Cloud. The

---

*Accepted for publication in Nature. Adapted with permission from *Neutrino Astrophysics: A Research Briefing*. Copyright 1995 by the National Academy of Sciences. Courtesy of the National Academy Press, Washington, D.C.

observed solar neutrino fluxes all agree with the theoretical expectations to within factors of two or three, successfully concluding a half-century struggle to understand how the sun shines. Nevertheless, persistent deficits of electron-type neutrinos exist in all four solar neutrino experiments, which has led to suggestions that neutrinos may have small masses and may oscillate between different types. New neutrino experiments are required to resolve the long-standing "solar neutrino problem." We shall also describe efforts that are underway to develop experiments that can detect astronomical neutrinos at high energies from sources that are well beyond the Galaxy. First, however, we shall sketch briefly the development and overall state of the field of neutrino astrophysics.

In the 1920's and 1930's, scientists developed the theoretical basis for understanding how the sun shines. They proposed[1,2] that the nuclear fusion reactions among light elements occur near the the center of the sun and provide the energy emitted by the sun for four and a half billion years. While most of this energy ends up as electromagnetic radiation from the surface, approximately three percent is believed to be emitted directly from the center of the sun in the form of neutrinos[3]. In a pioneering experiment that has been going on for 25 years[4], Raymond Davis, Jr. and his collaborators first detected solar neutrinos using radiochemical techniques and a cleaning fluid (perchloroethylene) as a target.

Over the last several years there has been striking confirmation of the astronomical predictions of solar neutrinos. The Kamiokande water Cerenkov detector located in the Japanese alps has identified solar neutrinos by their directionality, measured by observing the direction of recoil electrons from neutrino interactions[5]. Recently, two radiochemical experiments using gallium (the GALLEX experiment[6,7] in Italy and the SAGE experiment[8,9] in Russia) have detected the copious low energy (below 400 keV) neutrinos that are the primary component of the solar neutrino flux.

A supernova represents one of the most awesome events in astronomy. Type II supernovae are believed to result from the sudden gravitational collapse of a star that has completed cycles of nuclear burning. The energy calculated to be produced from this collapse is almost one thousand times larger than that observed as light. Theory indicates that more than 99 percent of the energy is emitted in the form of neutrinos[10]. In February 1987, neutrinos from SN1987a were detected as a series of pulses in two water Cerenkov detectors (Kamiokande in Japan[11] and IMB[12] in the United States), the neutrinos arriving, as expected, a few hours before the light from the explosion.

The definitive detection of solar neutrinos and the detection of neutrinos from SN1987a are two of the most remarkable scientific discoveries of the last decade. They provide dramatic confirmation of fundamental theories concerning stellar interiors. The detection of solar neutrinos demonstrates that fusion energy is the basic source of energy received from the sun. The observations of solar and supernova neutrinos open up a new area of science: neutrino astrophysics.

The observation of astrophysical neutrinos makes it possible to probe the innermost regions of stars, dense regions from which light cannot escape. Because of their very small interaction probabilities, neutrinos can escape from these inner regions; for the same reason, of course, it is difficult to detect these elusive particles.

The motivation for proposing[13,14] the first practical solar neutrino experiment was astrophysical, to test directly the hypothesis that stars shine and evolve because of nuclear fusion reactions in their interiors. However, it has become apparent in recent years that solar neutrinos provide a beam of elementary particles that can be used to investigate fundamental physics, in particular to study intrinsic neutrino properties.

Of particular importance is the question of whether neutrinos have mass and whether neutrinos transform (oscillate) from one type to another[15-18]. (There are believed to be three types of neutrinos: electron-type $\nu_e$, muon-type $\nu_\mu$, tau-type $\nu_\tau$. Only $\nu_e$ are created in the sun but all types are believed to be emitted from supernovae). The question of neutrino mass is of fundamental importance for particle physics and for cosmology. In the case of particle physics, neutrino mass may provide a clue to the problem of the origin of the masses of all particles. An interesting class of theories[19] called grand unified theories suggests that all interactions (strong, electromagnetic, and weak) are unified at a large energy scale and that the neutrino masses are inversely proportional to this scale. Two related ways to probe this very high energy scale are the search for proton decay and the search for very small neutrino masses.

Our basic ideas about cosmology that explain the microwave background radiation predict a similar background of relic neutrinos. If the heaviest of these neutrinos, usually taken to be $\nu_\tau$, has a mass above a few electron volts, these relic neutrinos could be an important component of dark matter. In this case, $\nu_\mu$ with a much smaller mass could play a decisive role for solar neutrinos.

The only available way to study very small neutrino masses is through the possibility of oscillations that transform one type of neutrino into another. For neutrino masses greater than $10^{-2}$ eV such oscillations have been looked for in experiments, not discussed in this paper, using neutrinos produced in the earth's atmosphere by cosmic rays or neutrinos produced in the laboratory by reactors or accelerators. For smaller neutrino masses ($\nu_\mu$ or $\nu_\tau$ with mass in the range $10^{-2}$ to $10^{-5}$) the only available possibility is to use solar neutrinos,[15,16] with oscillations depleting the flux of $\nu_e$ by converting some of them into $\nu_\mu$ or $\nu_\tau$. Of particular interest is the possibility that the oscillations may be enhanced[17,18] by the coherent interaction of the neutrinos with the solar material. This coherent interaction is called the MSW effect.

The observations made with the four experiments performed so far indicate fluxes of electron-neutrinos that are consistently below the results of astrophysical calculations. In fact, comparisons between the rates of different experiments suggest (independent of astrophysical uncertainties) that some previously unaccounted for physical process, perhaps involving non-zero neutrino mass, may be occurring. The experiments carried out so far are the first pioneering experiments and final conclusions require additional experiments that are currently being developed.

In this report, we also consider experiments designed to observe neutrinos of much higher energies, six to nine orders of magnitude larger than those of solar neutrinos. The potential sources of higher energy neutrinos, the techniques for their detection, and the scientific justifications for the research are all much different from the solar case. Unlike the sun, possible astrophysical sources of high-energy neutrinos are remote and not well understood; estimates of their neutrino fluxes are necessarily speculative. Detectors of high-energy neutrinos are complementary to detectors of solar neutrinos, being optimized for large areas rather than for low thresholds.

Candidates for sources of high-energy neutrinos include such exotic but relatively nearby objects as neutron stars and black holes in the Galaxy as well as the nuclei of distant but highly luminous galaxies. Since we observe cosmic rays too energetic to be confined within the Galaxy[20] we know that there exist enormously energetic processes that might well be sources of high energy neutrinos. Will any of these sources produce detectable numbers of high energy neutrinos? We will never know until we look.

## Ongoing solar-neutrino experiments

There is no single, best type of detector that can be used to observe solar neutrinos. The energies of the neutrinos vary over a range of one-and-half orders of magnitude and the fluxes from the most important neutrino sources vary by four orders of magnitude. According to standard solar model calculations, the numerous neutrinos from the initiating proton-proton reaction with energies less than 0.4 MeV, have a flux of about $6 \times 10^{10} \text{cm}^{-2}\text{s}^{-1}$; the rare, higher-energy neutrinos from $^8$B decay with energies up to 14 MeV have a flux of about $6 \times 10^6$ cm$^{-2}$s$^{-1}$. Eleven recently-published solar models, each constructed using a different computer code and different input physics, give fluxes for those neutrinos less energetic than 1 MeV which are the same to better than 10%[21]

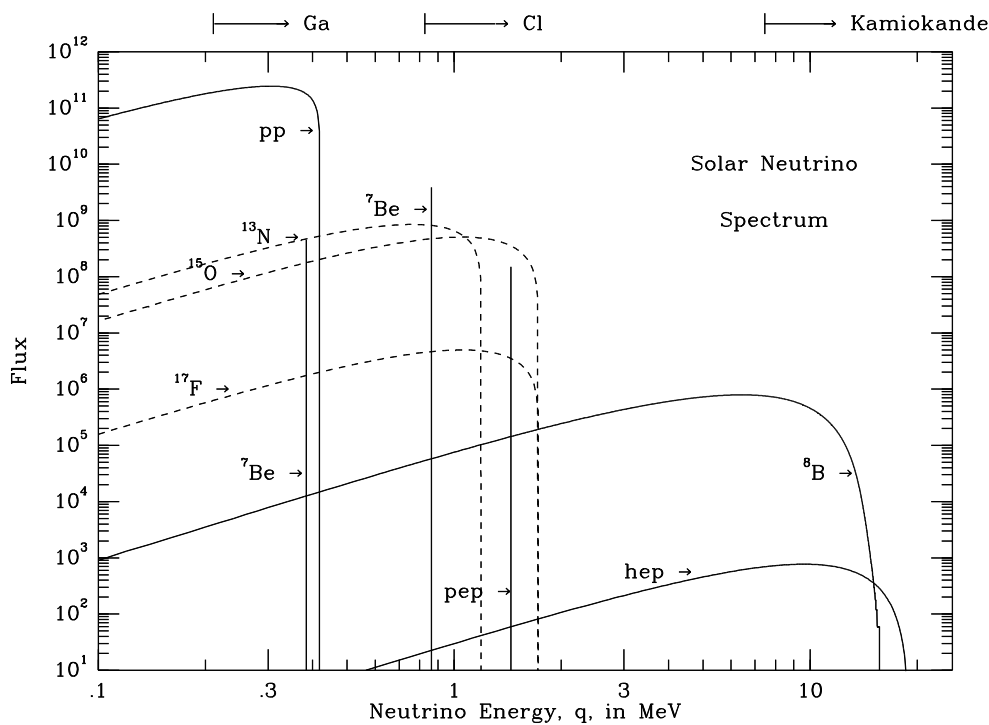

FIG. 1. Solar Neutrino Spectrum. This figure shows the energy spectrum of neutrinos predicted by the standard solar model[3]. The neutrino fluxes from continuum sources (like pp and $^8$B) are given in the units of number per cm$^2$ per second per MeV at one astronomical unit. The line fluxes (pep and $^7$Be) are given in number per cm$^2$ per second. The spectra from the pp chain are drawn with solid lines; the CNO spectra are drawn with dotted lines. The arrows at the top of the figure indicate the energy thresholds for the ongoing neutrino experiments.

Figure 1 shows the theoretically calculated solar neutrino spectrum from different neutrino sources. Because of the wide range in energies and in fluxes, a set of detectors with sufficient complementarity and some redundancy is required to untangle the different aspects of neutrino physics and of solar physics.

**The chlorine experiment.** In this remarkable experiment[4], neutrinos are detected using a large tank containing 615 tons of perchloroethylene, $C_2Cl_4$, which is a commercially available cleaning fluid. The target isotope, $^{37}Cl$, can capture a neutrino, producing a radioactive isotope $^{37}Ar$ by the following process:

$$\nu_e + {}^{37}Cl \rightarrow {}^{37}Ar + e^-. \qquad (1)$$

This reaction can occur for electron neutrinos with energy greater than 0.8 MeV. The neutrino reactions are registered by counting the radioactive nuclei, $^{37}Ar$. In order to avoid extraneous (background) interactions caused by cosmic rays incident on the surface of the earth, the experiment is conducted 1500 meters underground in the Homestake Gold Mine in Lead, South Dakota. After about two months in the solar-neutrino "sunshine", the buildup of $^{37}Ar$ atoms from neutrino capture by $^{37}Cl$ is approximately balanced by the loss due to decay. At this point, the standard solar model[22–24] predicts that only about 54 $^{37}Ar$ atoms are present in the 615 tons of $C_2Cl_4$. Over the past two decades, more than 100 extractions of $^{37}Ar$ have been performed. After correction for detection efficiency the average number of $^{37}Ar$ atoms in the detector at extraction is observed to be only 17, corresponding to a solar neutrino induced production rate of 0.5 atoms per day, far fewer than the 1.5 atoms per day expected on the basis of the standard model. In terms of the solar neutrino unit, SNU (pronounced 'snew', 1 SNU = $10^{-36}$ interactions per target atom per second), the observations yield $2.55 \pm 0.25$ SNU, about one third of the prediction of the standard model. For two decades, this discrepancy between the measured chlorine event rate and the event rate predicted by the standard model constituted the well-known Solar Neutrino Problem.

**The Kamiokande experiment**. Like a supersonic aircraft creating a sonic boom, a charged particle moving through matter faster than the speed of light in the matter gives off a conical shock wave of light called Cerenkov radiation. Sensitive photomultiplier tubes can record the flashes of light that betray the presence and direction of these high-speed particles. With the aid of special large photomultiplier tubes, experimenters working in a mine in the Japanese alps and using a 3000 ton detector of ultra-pure water (680 ton fiducial volume), called Kamiokande, were able to detect electrons that had been struck by neutrinos from the sun[5]. The reaction that was observed is

$$\nu + e^- \to \nu + e^- \ . \qquad (2)$$

At the energies of interest for solar neutrino detection, reaction (2) is about six times more likely for $\nu_e$ than for $\nu_\mu$ or $\nu_\tau$. Despite a significant background from radioactivity of the rocks in the mine, from radioactivity in the water, and from cosmic-ray induced events, the neutrinos were unambiguously detected because the electrons struck by neutrinos recoiled preferentially in the direction from the sun to the earth. Kamiokande, the first "active" (electronic rather than radiochemical) solar-neutrino experiment, showed two very important things: first, the detected neutrinos do indeed come from the sun, and, second, the observed flux above an electron-detection threshold of about 8 MeV is about half as large as predicted by the standard solar model. The discrepancy between prediction and observation is less severe at energies above 8 MeV (the Kamiokande experiment) than it is for all energies above 0.8 MeV(the chlorine experiment).

The predicted counting rate for any particular solar neutrino experiment depends upon calculations that are usually based upon a standard solar model. Solar models are constrained by measurements of the solar luminosity, age, chemical composition, and thousands of seismological frequencies. Nevertheless, the predicted neutrino emis-sion rates cannot be tested independent of solar neutrino experiments. Changes in the input data for the standard solar model could in principle be made just so as to fit any one of the existing experiments. However, it appears essentially impossible to explain simultaneously the results of the chlorine and the Kamiokande experiments by adjusting input solar data[23-25]. Just the $^8$B neutrinos that are observed in the Kamiokande experiment correspond to an expected counting rate in the chlorine experiment of $(3.21 \pm 0.46)$ SNU compared with the observed rate of $(2.55 \pm 0.25)$ SNU. However, in the chlorine experiment the solar model predicts an additional 1.2 SNU from the much more reliably calculated lower-energy neutrinos ($^7$Be and *pep* neutrinos, see Figure 1). This suggests that the lower-energy neutrinos might be preferentially depleted by some new weak interaction process (which could be the MSW effect discussed below).

**The gallium experiments**. The neutrinos produced by the most basic fusion reaction in the sun, the fusion of two protons, the so-called $p-p$ reaction, are too low in energy to be detected by either the chlorine or the Kamiokande experiment. Since the $p-p$ neutrinos are predicted to be the most numerous neutrinos emitted by the sun and since their flux is more reliably predicted than any other neutrino flux[3], the measurement of the terrestrial flux of $p-p$ neutrinos has long been seen as an essential goal of solar neutrino astronomy. The possibility of using gallium as a detector for these neutrinos was suggested almost thirty years ago[28] and a design for the experiment was developed[29]. Only in the last six years has it proved possible to carry out this important experiment.

The detector reaction is

$$\nu_e + {}^{71}\text{Ga} \rightarrow {}^{71}\text{Ge} + e^-, \qquad (3)$$

which has an energy threshold of 0.2 MeV. The gallium experiments operate similarly to the chlorine experiment, extracting and counting radioactive atoms of $^{71}$Ge. The standard model prediction is that after a month of operation about 16 atoms of $^{71}$Ge will be present in 30 tons of gallium.

One of the gallium experiments, GALLEX[6,7], is a European-American-Israeli collaboration operating in the Gran Sasso tunnel near Rome, Italy. GALLEX uses 30 tons of gallium in a GaCl$_3$-HCl solution. The other experiment, SAGE[8,9] (the Soviet-American Gallium Experiment) is operating in a subterranean laboratory under Mount Andyrchi of the Caucasus Mountains in the southern region of Russia. SAGE currently uses 60 tons of molten gallium metal.

The fact that the detectors use gallium in two very different chemical environments, the GaCl$_3$-HCl solution (GALLEX) and the metallic form (SAGE), provides a consistency check on the results. The current Gallex results are $79 \pm 12$ SNU and the current SAGE results are $74 \pm 14$ SNU. These results also are significantly below the standard model prediction, which is $132 \pm 7$ SNU for gallium detectors. The GALLEX detector response to neutrinos has been experimentally checked[30] using a source ($^{51}$Cr) of 1.7 Megacurie that emits neutrinos of 750 keV and 430 keV from electron capture. A similar check is planned for the SAGE detector.

**Summary**. Thus as time has passed and more experiments have been performed, the solar neutrino problem has not gone away. Indeed, it appears that, independent of detailed solar model calculations, there is no combination of the expected neutrino fluxes that fits all the available data.

On the other hand, the results of all four of the experiments (chlorine, Kamiokande, GALLEX, and SAGE) can be explained[31,32] if either $\nu_\mu$ or $\nu_\tau$ has a mass of about 0.003 eV and has a small amount of mixing with a lighter $\nu_e$. In this case the $\nu_e$ produced at the center of the sun are transformed (with a transformation probability that depends upon the energy of $\nu_e$) into $\nu_\mu$ or $\nu_\tau$ as they pass through the material medium of the sun (the MSW effect[17,18]). Only future experiments can determine if this is indeed the correct explanation.

## Solar-neutrino experiments under construction

Solar neutrino detectors fall into two classes, "active" (like Kamiokande) and "radiochemical" (like chlorine and gallium). Active detectors provide information about the arrival time, direction and energy of individual neutrinos, and in some cases the neutrino type (e.g., electron type or muon type), but are complex and often faced with persistent backgrounds. Radiochemical detectors are sensitive only to electron-type neutrinos (i. e., they specify uniquely the neutrino type) and provide a single number for the rate of detection of all neutrinos above an energy threshold, a number that is averaged over a period of time comparable to the mean life (typically weeks or months) of the radioisotope that is produced.

In this section, we describe the three active and one radiochemical solar neutrino detectors that are under construction. All of these detectors will have much higher event rates, for the active detectors several thousand events per year. The expected high event rates will make it possible to search for time dependence in the neutrino flux, in particular the expected seasonal variation due to the orbital eccentricity of the earth ($\sim 7\%$ effect, peak-to-peak).

**The Sudbury Neutrino Observatory: SNO.** Deuterium ("heavy" hydrogen, $^2H$, with a neutron and a proton in its nucleus) is an excellent target for neutrinos. Two neutrino interactions can occur[33]:

$$\nu_e + {}^2H \to p + p + e^- \; ; \tag{4a}$$

$$\nu_x + {}^2H \to p + n + \nu_x \; . \tag{4b}$$

Reaction (4a) can only be induced by electron-type neutrinos, whereas the cross-section for reaction (4b) is the same neutrinos of all three types. If more neutrinos are detected via reaction (4b) than by reaction (4a), that would be direct evidence that some electron-type neutrinos have oscillated into neutrinos of some other type.

The SNO experiment is located near Sudbury, Ontario in one of the deepest mines in the Western hemisphere (a nickel mine belonging to the INCO company). A large cavern has been excavated 2070 m underground to hold a detector consisting of 1000 tonnes of heavy water. When SNO begins operation in 1996, solar neutrinos are expected to be detected at the rate of more than 10 counts per day via reactions (4a) and (4b). In addition to the unique reactions of neutrinos on deuterium, SNO will observe the same neutrino-electron scattering process that is detected by Kamiokande.

If electron-type neutrinos oscillate into one of the other known neutrino types as they travel from the interior of the sun to the terrestrial detector, this will be revealed by the comparison of the rates in the two deuterium reactions, and, if the oscillation parameters are favorable, also by its distinctive effect on the shape of the observed electron energy spectrum in reaction (4a). These potential signatures of new physics are independent of solar models and solar physics.

**Superkamiokande.** The Superkamiokande experiment[34], a much improved version of the current Kamiokande experiment, utilizes 50,000 tonnes of ultra-pure ordinary water. This immense detector will, beginning in 1996, provide a thirty-fold increase in the observed rate of neutrino-electron scattering events. The change in the shape of the neutrino energy spectrum predicted by the MSW effect may be observable in this experiment via the spectrum of recoil energies of the scattered electrons.

Both Superkamiokande and SNO detect only neutrinos above 5 MeV.

**BOREXINO.** BOREXINO[35] will be the first active detector with the goal of observing the intense flux of low-energy neutrinos produced by electron capture on beryllium nuclei (see the 0.86 MeV $^7$Be neutrino line in Figure 1). The BOREXINO detector is to be constructed with standard liquid scintillator materials, except that the requirement for radiopurity (uranium and thorium less than $\sim 10^{-16}$ g/g) is unprecedently stringent. If the MSW interpretation is correct, then the observed flux of $^7$Be neutrinos will be greatly reduced relative to the predictions of the standard solar model.

**IODINE**. An iodine radiochemical detector (all $^{127}$I) will be somewhat similar to the existing chlorine detector but, per target atom, is expected to have a higher counting rate[36]. A modular 100-ton iodine detector is now under construction in the Homestake Mine near the present chlorine detector and will begin operating in mid-1995.

Calibration experiments will be performed to measure the probability for neutrino-induced conversion of $^{127}$I to $^{127}$Xe. These experiments are necessary to determine the relative sensitivity of the detector to $^7$Be and $^8$B neutrinos and to interpret the total rate. Expansion of the present iodine detector may be proposed after operating experience is obtained with the current detector and after the neutrino cross sections are measured.

**Developing new solar neutrino detectors** Additional solar neutrino detectors are required to obtain more complete information on the different portions of the solar neutrino spectrum.

The 5000 ton ICARUS detector[37] will be installed in the Gran Sasso laboratory and is planned to be operational in 1998. Like the chlorine detector, ICARUS is based on an inverse beta process, $\nu_e + ^{40}Ar \to e^- + ^{40}K^*$, but operates in real time and is sensitive only to $^8$B neutrinos. The detector uses liquid argon with the goal of measuring the recoil electrons and the decay gamma rays (from the $^{40}K$ excited state) that are produced by neutrino capture.

A major future goal is to measure the energy and arrival time of the lowest energy neutrinos, the *pp* neutrinos which are detected by the radiochemical method in the gallium experiments. The proposed HERON detector[38] is based on the ballistic propagation of rotons (sound excitations) in liquid helium in the superfluid state with detection of the heat pulses by an array of bolometers. The goal is to achieve a threshold of the order of 10 keV, suitable for measuring the recoil electron spectrum and rate of neutrino-electron scattering from both $pp$ and $^7Be$ neutrinos. The proposed HELLAZ detector[39] also would detect neutrino-electron scattering in helium, but in high-pressure low-temperature gas.

## High-energy neutrinos

Higher-energy neutrinos that are most accessible experimentally have energies above a $TeV$ ($10^{12}$ $eV$ = $10^6$ MeV). Astrophysical neutrinos with these energies can be produced by high-energy collisions among nuclei or between protons and photons that produce secondary particles (pions and muons), which have neutrinos among their decay products. Such collisions are believed to take place when matter falls into a neutron star from a companion star in a double star system. Also such collisions are expected on a large scale at the centers of active galactic nuclei (AGN). However the flux of high energy neutrinos[40,41] from such sources cannot be calculated with accuracy. It is only by searching for these neutrinos that we can probe the processes taking place deep in the interior of these (and other) astrophysical systems.

High-energy neutrinos can be detected via the reaction of a muon neutrino with a nuclear particle, leading to the neutrino being transformed into a charged muon. A multi-$TeV$ charged muon will travel many kilometers in earth or in water. The neutrino target volume is thus the area of the muon detector times the range of the muon. Target volumes of billions of tons may be achieved this way. The fast-moving muons can be detected by observing the blue flash of Cerenkov light (see discussion of Kamiokande experiment) they produce when traversing a transparent medium, specifically water or ice. Detectors with effective areas of a few hundred square meters have been built in mines, four new instruments are being built with areas in the 20,000 m² range, and plans are being discussed for one or more instruments in the 1 km² class[42].

The Kamiokande and the IMB detectors have observed neutrinos of $10^8$ to $10^{11}$ eV produced by cosmic rays in the atmosphere, but no neutrinos from sources outside the solar system have been observed with the exception of the observations of SN1987A. Estimates[40] of the potential fluxes of high energy neutrinos from sources such as active galactic nuclei suggest that a surface area greater than 0.1 km² is necessary for meaningful observations. Furthermore, since the expected signal-to-noise improves greatly with the increasing threshold up to TeV energies, detectors of great thickness are desirable. It is expected generally agreed these requirements can only be met by the use of large bodies of water such as lakes, the ocean, or polar ice.

The first instruments designed specifically for high energy neutrino detection are now under construction in Siberia in Lake Baikal ([43]), off Hawaii in the deep ocean (DUMAND[44]), at the South Pole in the deep polar ice (AMANDA[41]), and in the Mediterranean near Pylos, Greece (NESTOR[45]). All four of the instruments being built aim at detection areas of the order of 20,000 m² and an angular resolution of order 1° for TeV muons.

It is expected that high-energy neutrino astronomy will require a telescope with an effective muon detecting area of at least 1 km by 1 km. Two of the principal science goals for a kilometer-scale neutrino telescope are the search for neutrinos from the annihilation in the sun of weakly interacting dark matter particles (with masses in the GeV-TeV range) and the observation of individual active galactic nuclei with good statistics. Studies of neutrino oscillations may also be possible[40] The four instruments presently under construction represent an increase of about a factor of 50 in area from previous mine-based detectors; another increase of 50 is needed to reach the desired scale.

## What next?

Solar-neutrino experiments carried out over the past thirty years have closed an important chapter in the history of science that began in the early part of this century. We now have conclusive, direct evidence for nuclear-fusion reactions that occur among light elements in the center of the sun. At the same time the sun provides a unique source of neutrinos for studying the possibility of very small neutrino masses.

Solar-neutrino experiments therefore tell us both about elementary-particle physics and about the details of nuclear fusion within the innermost region of the nearest star.

Experiments that are currently underway have the potential to establish–independent of theoretical calculations carried out with solar models–whether new physics is required to explain the solar neutrino observations. If the number of neutrinos that are observed by the Sudbury Neutrino Observatory (SNO) in reaction (4b), which can be induced by any of the known types of neutrinos, exceeds the number observed in reaction (4a), which can only be initiated by electron-type neutrinos, this will be direct evidence for neutrino oscillations. The measurements of a neutrino energy spectrum by the SNO and the Superkamiokande experiments will test whether the spectrum of the higher energy ($^8$B) solar neutrinos that arrive at earth from the sun is the same as the energy spectrum of neutrinos from the same radioactive isotope observed in the laboratory or has been distorted by neutrino oscillations. However it is possible that even if oscillations take place their effect on the $^8B$ neutrinos, the only ones detected by SNO and Superkamiokande, may be too small to be detected. Fortunately, theoretical calculations of the MSW effect imply that the deficit of electron-type neutrinos is much larger for the lower-energy $^7$Be neutrinos; if this is correct, the flux detected in BOREXINO will be much smaller than predicted by the standard model.

Solar neutrino experiments are fundamental both for physics and for astronomy. It is essential that we not rely on just a few experiments, since the history of science has shown that systematic uncertainties can sometimes lead to mistaken conclusions unless results are checked by performing measurements in different ways. This is particularly important when the measurements are as intrinsically difficult as they appear to be for solar neutrino experiments.

The timing of SN1987a was extremely fortunate since the water Cerenkov detectors had only been operating for a few years. An important lesson of that supernova neutrino detection is that neutrinos could reveal supernovae in our galaxy even if the light is obscured by the galaxy itself. It would be very desirable that a number of detectors are prepared to detect neutrinos from the next supernova in our galaxy.

The last fifty years have seen the extension of astronomy from the optical spectrum to the whole range of the electromagnetic spectrum from radio waves to gamma rays. In that process, many new aspects of the astronomical world have been uncovered. Neutrino astronomy can provide a unique window by means of which one can see into the deepest interiors of stars and galaxies. So far the only sources detected are the sun and supernova 1987a. Neutrino astronomy is an exciting frontier for exploration in the twenty-first century.

## Acknowledgements

We are grateful to many of our colleagues who have acted as informal referees of individual sections of this report and who have worked hard to help us make the material as accurate as we could. We are especially grateful to Dr. R. Riemer, Associate Director, Board on Physics and Astronomy, National Research Council, and to Dr. D. Schramm, Chair, Board on Physics and Astronomy, NRC, for wise counsel and for their leadership in guiding this review.